%% file: bare_jrnl_new_sample4.tex
\documentclass[lettersize,journal]{IEEEtran}
\usepackage{amsmath,amsfonts,amssymb}
\usepackage{algorithmic}
\usepackage{algorithm}
\usepackage{array}
\usepackage{caption}
\usepackage{subcaption}
\usepackage{textcomp}
\usepackage{stfloats}
\usepackage{url}
\usepackage{xcolor}
\usepackage{verbatim}
\usepackage{graphicx}
\usepackage{cite}

\usepackage{bm}
\usepackage{bbm}
\usepackage{stfloats}
\usepackage{verbatim}
\usepackage{float}
\usepackage{array}
\usepackage{rotating}
\usepackage{gensymb}
\usepackage{comment}
\usepackage{multirow}
\usepackage{lscape}
\usepackage{hhline}
\usepackage{makecell}
\usepackage[inline]{enumitem}

\input{commands}

\begin{document}

\title{Full-Duplex V2X Integrated Sensing and Communication Scenario: Stochastic geometry, Monte-Carlo, and Ray-Tracing Comparison}

\author{François De Saint Moulin\IEEEauthorrefmark{1}, Simon Demey, Charles Wiame\IEEEauthorrefmark{2}, Luc Vandendorpe\IEEEauthorrefmark{3}, Claude Oestges\IEEEauthorrefmark{4}
\\
\IEEEauthorblockA{ICTEAM, UCLouvain - Louvain-la-Neuve, Belgium\\}
\IEEEauthorrefmark{1}francois.desaintmoulin@uclouvain.be,
\IEEEauthorrefmark{2}charles.wiame@uclouvain.be,
\IEEEauthorrefmark{3}luc.vandendorpe@uclouvain.be,
\IEEEauthorrefmark{4}claude.oestges@uclouvain.be
\thanks{François De Saint Moulin is a Research Fellow of the Fonds de la Recherche Scientifique - FNRS.}}

\maketitle

\begin{abstract}
In this paper, performance of an Integrated Sensing and Communication (ISAC) Vehicle-to-Everything (V2X) scenario is evaluated, in which a vehicle simultaneously detects the next vehicle ahead while receiving a communication signal from a RoadSide Unit (RSU) of the infrastructure. Univariate and joint radar and communication performance metrics are evaluated within three different frameworks, namely the Stochastic Geometry (SG), Monte-Carlo (MC), and Ray-Tracing (RT) frameworks. The parameters of the system model are extracted from the RT simulations, and the metrics are compared to assess the accuracy of the SG framework. It is shown that the SG and MC system models are relevant w.r.t. RT simulations for the evaluation of univariate communication and sensing metrics, but larger discrepancies are observed for the joint metrics.
\end{abstract}


\section{Introduction}
\label{sec:introduction} 
\input{introduction}

\section{System Model}
\label{sec:system_model} 
\input{system_model}

\section{Performance Metrics}
\label{sec:performance_metrics} 
\input{metrics}

\section{Parameters Fitting}
\label{sec:parameters_fitting}
\input{parameters_fitting}

\section{Numerical Analysis}
\label{sec:numerical_analysis}
\input{numerical_analysis}

\section{Conclusion}
\label{sec:conclusion}
\input{conclusion}

\bibliographystyle{IEEEtran}
\bibliography{biblio}

\end{document}

%% file: commands.tex
\newcommand{\V}{{\text{V}}}
\renewcommand{\L}{{\text{L}}}

\renewcommand{\t}{{\text{t}}}
\renewcommand{\r}{{\text{r}}}

\newcommand{\R}{\text{R}}
\newcommand{\C}{\text{C}}
\newcommand{\I}{\text{I}}

\newcommand{\ones}[1]{\mathbbm{1}\left(#1\right)}

%% file: introduction.tex
In future networks, V2X communications are crucial for making vehicles smarter, and potentially autonomous. Extensive research is being conducted in these networks as they have the potential to enhance the safety and driving experience of road users. However, high accuracy localization and high-throughput communication are required. The use of ISAC systems may therefore help, as it addresses spectrum congestion issues, reduces hardware requirements on V2X network nodes, and potentially improves performance for both functions, possibly with cooperation. Nevertheless, the performance of both functions in such networks must be evaluated accurately to effectively design these systems.\smallskip 

For that extent, multiple frameworks have been proposed. A first option is the SG framework \cite{9378781}. In that case, the nodes positions are abstracted as random point processes, enabling to obtain closed-form expressions for average performance metrics. Using this framework, many automotive scenarios have been studied, e.g. \cite{9119440,7819520,9266343,9124830,8967012}. The most common alternative is the MC framework, often used to validate the expressions obtained with SG. Instead of obtained closed-form expressions, the performance metrics are evaluated through numerous numerical  simulations of the scenario. All the networks statistics are obtained, but it is computationally expensive. A third option is the RT framework \cite{mcnamara1990introduction}, In that case, the scene is modelled with a high level of details, and the transmission is modelled through the propagation of rays \cite{7152831}. This framework is assumed to be the most accurate, but also the most computationally expensive. \smallskip 

In this paper, the results developed in \cite{consanguinite} within the SG framework are extended by comparing the proposed univariate and joint metrics for the evaluation of radar and communication performance with the three different frameworks. The same V2X scenario is considered, in which a vehicle simultaneously detects the next vehicle driving ahead, while receiving a communication signal from the infrastructure. First, The parameters of the three frameworks are extracted from the RT dataset. Then, univariate and joint ISAC metrics are evaluated and compared.\smallskip

In this paper, the system model used with SG and the particularities introduced for the MC and RT frameworks are first presented in Section \ref{sec:system_model}. Then, the evaluated radar and communication metrics are presented in Section \ref{sec:performance_metrics}. Next, the parameters fitting is detailed in Section \ref{sec:parameters_fitting}. Finally, the metrics are compared with the three frameworks in Section \ref{sec:numerical_analysis}.

%% file: system_model.tex
\begin{figure}
\centering 
\includegraphics[width=\linewidth]{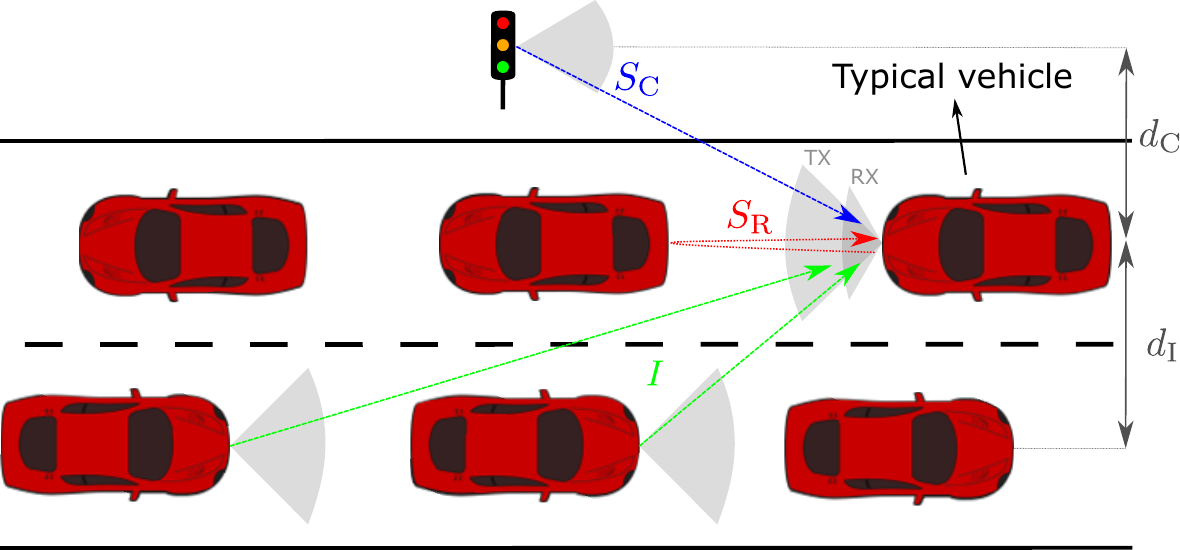}
\caption{V2X ISAC scenario.}
\label{fig:V2X_ISAC_scenario}
\end{figure}

Let us consider the two-lane automotive scenario illustrated in Figure \ref{fig:V2X_ISAC_scenario}. The vehicles are all equipped at the front side with an ISAC system. The transmitted powers, and the antenna beamwidths at the transmitter and receiver of the vehicles are respectively denoted as $P_\V$, $\phi_{\V \t}$, and $\phi_{\V \r}$.  The typical vehicle receives simultaneously a radar echo reflected by the first vehicle in front and driving in the same lane, and a communication signal from the nearest RSU ahead. The transmitted power and antenna beamwidth of the RSU at the transmitter are respectively denoted as $P_\L$ and $\phi_{\L \t}$. The distance perpendicular to the road direction between the typical vehicle and the RSU (resp. the centre of the opposite lane) is denoted $d_\C$ (resp. $d_\I$). The minimum and maximum detectable range of the radar function, the minimum distance parallel to the road of the communication signal, and the minimum distance parallel to the road of the interfering signals, are respectively denoted as $r_{\R\text{min}}$, $r_{\R\text{max}}$, $r_{\C\text{min}}$, and $r_{\I\text{min}}$. Note that $r_{\C\text{min}}$ and $r_{\I\text{min}}$ are directly related to the antenna beamwidths of the different systems. It is assumed that this scenario is interference-limited, owing to the high interference arising from the vehicles driving in the opposite direction, and the interference from the other function.\smallskip

Full-duplex ISAC is supposed in this automotive application. Namely, for the communication function, the radar echo generated by the typical vehicle is considered as an interference. Contrariwise, for the radar function, the communication signal transmitted by the RSU is considered as an interference. Moreover, the signals transmitted by the vehicles driving in the opposite lane are interfering with both functions. The self-interference is not considered in this work, as this issue is not specific to ISAC scenarios, but to full-duplex systems in general. It is assumed that it has been mitigated through electromagnetic isolation, beamforming, self-interference cancellation, or other techniques \cite{10159012}.

\subsection{Nodes Distribution}
Assuming that the antennas of the typical vehicle are located at $r = 0$ m, the vehicles driving in the direction of the typical vehicle, the interfering vehicles on the opposite way, and the RSUs are distributed on three parallel lines following three independent Poisson Point Processes (PPP) $\boldsymbol{\Phi}_\R$, $\boldsymbol{\Phi}_\C$, and $\boldsymbol{\Phi}_\I$, of intensities respectively given by 
\begin{align}
    \lambda_\R(r) &= \lambda_\R \: \ones{r_{\R\text{min}} \leq r \leq r_{\R\text{max}}}, \\
    \lambda_\C(r) &= \lambda_\C \: \ones{r_{\C\text{min}} \leq r}, \\
    \lambda_\I(r) &= \lambda_\I \: \ones{r_{\I\text{min}} \leq r},
\end{align}
where $\lambda_\R$, $\lambda_\I$ and $\lambda_\C$ are respectively the densities of the vehicles driving on the same lane as the typical vehicle, the density of interfering vehicles, and the density of RSUs. PPPs have been selected to model all the node locations, owing to their tractability, and analytical flexibility. The integration of such point processes in this framework is discussed in \cite{consanguinite}.

\subsection{Propagation Models}
Let us consider a linear path-loss model written as $[L_k(r)]_\text{dB} = [\beta_k]_\text{dB} + 10\:\alpha_k \: \log_{10} r$, where $k \in \{R,C,I\}$ designates the considered link, $\alpha_k$ is the path-loss exponent, and $\beta_k$ the intercept. For the communication link, the received power $S_\C$ is given by the Friis formula:
\begin{equation}
S_\C(r_\C) = \rho_\C \: \left(r_\C^2 + d_\C^2\right)^{-\frac{\alpha_\C}{2}},
\end{equation}
with $\rho_\C = P_\L G_\C c^2 / 4\pi f_c^2 \beta_\C$. In this equation, $r_\C$ denotes the distance between the two nodes parallel to the road, $f_c$ is the carrier frequency, and $G_\C$ is the gain of the complete communication link. The small-scale fading is neglected for this link within the SG framework, for the sake of tractability. However, it is assumed to be highly Ricean, since the RSU is located on the same side of the road than the typical vehicle, and high frequencies (i.e. 24 or 77 GHz) are considered in automotive scenarios.\smallskip

For the interfering links, the total interfering power is also computed with the Friis formula: 
\begin{equation}
I = \sum_{i | x_i \in \boldsymbol{\Phi}_\I} \rho_\I \: \left(x_i^2 + d_\I^2\right)^{-\frac{\alpha_\I}{2}} \: |h_i|^2,
\end{equation}
where $\rho_\I = P_\V G_\I c^2 / 4 \pi f_c^2 \beta_\I$. The beamforming gain of the complete interfering link is denoted as $G_\text{I}$, and $|h_i|^2$ are small-scale fading random variables, supposed to be Ricean-distributed. \smallskip

Finally, for the radar link, the received power $S_\R$ is computed following a GO model, as used in ray-tracing applications \cite{7152831}:
\begin{equation}
S_\R(r_\R) = \rho_\R \: r_\R^{-\alpha_\R},
\end{equation}
where $\rho_\R = P_\V G_\R c^2 / 4\pi f_c^2 \beta_\R$. In this equation $r_\R$ denotes the distance parallel to the road between the antennas and the next vehicle ahead. The gain of the complete radar link is denoted as $G_\R$. The small-scale fading is also neglected, as it is again supposed to be highly Ricean. The GO model is also motivated by the high carrier frequencies considered in automotive scenarios.

\subsection{Ray-Tracing System Model}
\label{sec:RT_system_model}

\begin{figure}
\centering 
\includegraphics[width=\linewidth]{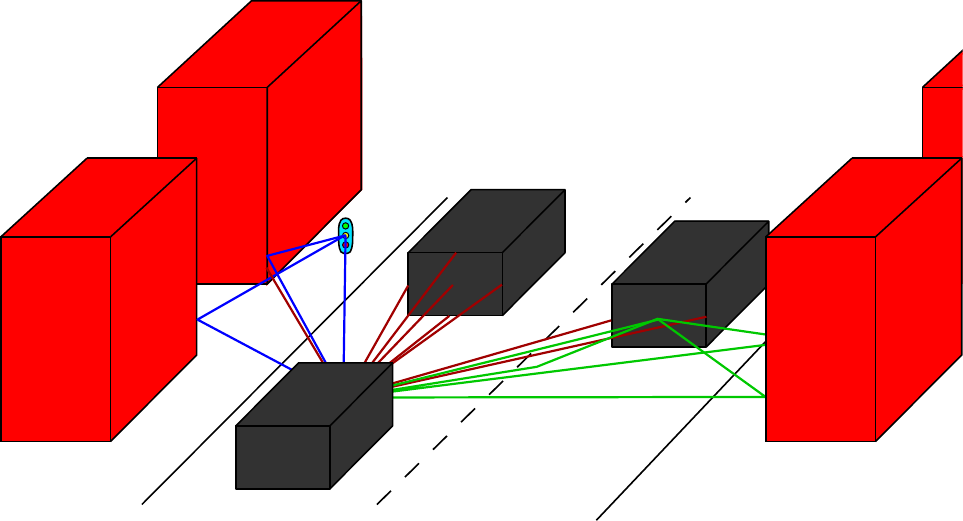}
\caption{Illustration of a ray-tracing simulation for the V2X ISAC scenario. The communication, radar, and interfering paths are respectively drawn in blue, red and green.}
\label{fig:RT_scenario}
\end{figure}

With RT, all the paths comprising up to two interactions with the environment are considered. Diffraction on the edges of the buildings and vehicles is only considered at the last interaction. The following differences may be highlighted in the generation of the environment:
\begin{enumerate}[label=(\roman*)]
  \item buildings are modelled along the streets as rectangular blocks. Perpendicular streets are generated at coordinates following another PPP. This generates additional multipath in the RT simulations. Thus, for the radar link, these should be filtered out to isolate the useful radar echo, arising from the next vehicle ahead, w.r.t. the other radar echoes from the environment. This is further discussed in Section \ref{sec:parameters_fitting};
  \item vehicles are modelled as rectangular blocks. Stacking vehicles are removed from the street. Thus, the position of the vehicles is better modelled by HCPPs intead of PPPs. Still, such PPs are often intractable, and the PPP is a good approximation if the nodes density is adapted accordingly \cite{5934671};
  \item vehicles are positioned in both lanes on the street following two PPPs with the same density. The interfering vehicles are then selected by performing a thinning with a given probability of interference $p_\I$, selected such that $\lambda_\I = p_\I \lambda_\R$. Therefore, the vehicles in the opposite lane are not all interfering, but they may still act as obstacles;
  \item The transmit and receive antennas are located at non-zero heights. For the ISAC systems embedded on the vehicles, they are both located at the front, but slightly separated from each other.
\end{enumerate}
Figure \ref{fig:RT_scenario} illustrates a RT simulation for the V2X ISAC scenario. One may notice that multipath is present for every link, and the radar function receives additional echoes from the environment. The RT parameters are summarised in Table \ref{tab:RT_parameters}.

\begin{table}
  \begin{center} 
  \caption{Parameters of the RT simulations.}
  \renewcommand{\arraystretch}{1.2}
  \begin{tabular}{|l|c|}
      \hline
      \multicolumn{1}{|c|}{\textbf{Parameters}} & \textbf{Values} \\\hline\hline
      3D position of the radar TX antenna & $(6.1,16,0.5)$ m \\\hline
      3D position of the radar RX antenna & $(6.1,16.2,0.5)$ m \\\hline
      3D position of the RSU TX antenna & $(6.1+r_\C,18,2.5)$ m \\\hline
      3D position of the interfering TX antennas & $(6.1+x_i,13,0.5)$ m \\\hline
      Streets length & 2 km \\\hline
      Streets width & 9 m \\\hline
      Perpendicular streets density & 10 km$^{-1}$ \\\hline
      RSU density $\lambda_\C$ & 10 km$^{-1}$ \\\hline
      Vehicles density $\lambda_\R$ & 20 km$^{-1}$ \\\hline
      Interfering vehicles density $\lambda_\I$ & 2 km$^{-1}$ \\\hline
      Carrier frequency $f_c$ & 26 GHz \\\hline
      Transmitted powers $P_\V$ and $P_\L$ & 20 dBm \\\hline
      Beamforming gains $G_\R$, $G_\C$ and $G_\I$ & 1 \\\hline
      Antennas beamwidths $\phi_{\text{V}\text{t}}$, $\phi_{\text{Vr}}$ and $\phi_\text{Lt}$ & \Gape[1pt][1pt]{\makecell{$\phi_{\text{V}\text{t}}=22.5\degree$ \\$\phi_{\text{Vr}}=\phi_\text{Lt}=45\degree$}} \\\hline
      Minimum radar detectable range $r_{\R\min}$ & 5 m \\\hline
  \end{tabular}
  \renewcommand{\arraystretch}{1.}
  \label{tab:RT_parameters}
  \end{center}
The distance $r_\C$ denotes the distance between the receive antenna of the vehicle, and the first RSU along the street. The positions $x_i \in \boldsymbol{\Phi}_\I$ define the positions of the interfering antennas along the other lane.
\end{table}

\subsection{Monte-Carlo System Model}
\label{sec:MC_system_model}

To further improve the propagation models considered with the SG, two additional features are added: 
\begin{enumerate}[label=(\roman*)]
\item Line-of-Sight (LoS) and Non Line-of-Sight (NLoS) links are considered. Owing to the presence of numerous vehicles on both lanes, and buildings acting as obstacles, both LoS and NLoS propagations occur for the communication and interfering links. Therefore, for these links, propagation parameters are defined for both LoS or NLoS links. The probability for the considered link to be in LoS is denoted as $p_{\text{L},v}(r)$, with $v \in \{\C,\I\}$, and $r$ the distance between the two nodes. In order to determine which nodes are in LoS or NLoS, independent thinnings are performed on the PPPs of the RSUs and interfering vehicles with this probability. The LoS probability model considered in this chapter is the 3GPP $d_1/d_2$ model \cite{channeltxt1}:
\begin{equation}
  p_{\text{L},v}(r) = \text{min}\left(\frac{d_{1,v}}{r},1\right)\left[1-e^{-\frac{r}{d_{2,v}}}\right] + e^{-\frac{r}{d_{2,v}}}, \label{eq:LoS_prob}
\end{equation}
with $d_{1,v}$ and $d_{2,v}$ being two fitting parameters for the two types of link;
\item Small-scale fading is considered for every link. Even if it is supposed to be highly Ricean for the radar and communication links within the SG framework, the introduction of such random variables helps to improve the accuracy of the fitting with the RT simulations.
\end{enumerate}
Since these features affects the mathematical tractability of the SG, they are only considered when MC simulations are performed.

%% file: metrics.tex
\subsection{Univariate Metrics}

On the one hand, for the communication function, the \textit{coverage probability} analysed. It is defined as the probability for the Signal to Interference-plus-Noise Ratio (SINR) of the communication function to be sufficient w.r.t. a given SINR threshold $\eta_\C$. For an interference-limited scenario, it is written as 
\begin{equation}
\mathcal{C}(\eta_\C) \triangleq \mathbb{P}\left(\frac{S_\C}{S_\R + I} \geq \eta_\C\right).
\end{equation}

On the other hand, for the radar function, the radar \textit{detection} and \textit{coverage probabilities} are considered. The former is defined as the probability for the SINR of the radar function to be sufficient w.r.t. a given SINR threshold $\eta_\R$. The latter is defined as the probability for the total received power at the radar function to be higher than a given detection threshold $\gamma_\R$. For an interference-limited scenario, they are respectively written as
\begin{align}
\mathcal{S}(\eta_\R) &\triangleq \mathbb{P}\left(\frac{S_\R}{S_\C + I} \geq \eta_\R\right), \\
\mathcal{D}(\gamma_\R) &\triangleq \mathbb{P}\left(S_\R + S_\C + I \geq \gamma_\R\right).
\end{align}
The detection probability is usually evaluated together with the \textit{false alarm probability}. However, the latter is not considered in this work, as the metrics are compared with each other to assess the accuracy of the SG framework, and the overall performance are not analysed.

\subsection{Joint Metrics}
In order to evaluate jointly the performance of both radar and communication functions, the following joint metrics have been defined: 
\begin{enumerate}
\item the \textit{Joint Radar Detection and Communication Coverage Probability} (JRDCCP) is the probability to detect a target at the radar function, while achieving a sufficient SINR at the communication function. It is defined from the detection and coverage probabilities as 
\begin{equation}
\mathcal{J}_\text{D}(\eta_\C,\gamma_\R) \triangleq \mathbb{P}\left(S_\R + S_\C + I \geq \gamma_\R, \frac{S_\C}{S_\R + I} \geq \eta_\C\right);
\end{equation}
\item the \textit{Joint Radar Success and Communication Coverage Probability} (JRSCCP) is the probability to achieve simultaneously sufficient SINRs at both radar and communication functions. It is defined from the success and coverage probabilities as
\begin{equation}
\mathcal{J}_\text{S}(\eta_\C,\eta_\R) \triangleq \mathbb{P}\left(\frac{S_\R}{S_\C + I} \geq \eta_\R, \frac{S_\C}{S_\R + I} \geq \eta_\C\right).
\end{equation}
\end{enumerate}

%% file: parameters_fitting.tex
\begin{table}
    \begin{center}
    \caption{Propagation parameters obtained with LSE fitting.}
    \renewcommand{\arraystretch}{1.2}
    \begin{tabular}{|c|c||c|c|c|}
    \cline{3-5}\noalign{\vskip-0pt}
    \multicolumn{2}{c|}{\rule{0pt}{9pt}} & \textbf{Mixed} & \textbf{LoS} & \textbf{NLoS} \\\hhline{--===}
    \multirow{2}{*}{\textbf{Radar}} & $\hat{\alpha}_\R$ & 1.74 & --- & --- \\\cline{2-5}
    & \Gape[2pt][1pt]{$\left[\hat{\beta}_\R\right]_\text{dB}$} & 3.72 & --- & --- \\\hline
    \multirow{2}{*}{\textbf{Comm.}} & $\hat{\alpha}_\C$ & 2.20 & 1.53 & 2.32 \\\cline{2-5}
    & \Gape[2pt][1pt]{$\left[\hat{\beta}_\C\right]_\text{dB}$} & 0.42 & 3.37 & 0.40 \\\hline
    \multirow{2}{*}{\textbf{Interference}} & $\hat{\alpha}_\I$ & 4.64 & 1.12 & 4.35 \\\cline{2-5}
    & \Gape[2pt][1pt]{$\left[\hat{\beta}_\I\right]_\text{dB}$} & 0 & 14.41 & 0 \\\hline
    \end{tabular}
    \renewcommand{\arraystretch}{1}
    \label{tab:RT_prop_params}
\end{center}
The parameters of the radar links have been computed considering only the power of the radar echo of the next vehicle ahead, while the environment echoes are filtered out of the dataset.
\end{table}

\begin{figure}
\centering 
\begin{subfigure}{\linewidth}
    \centering
    \includegraphics[width=\linewidth]{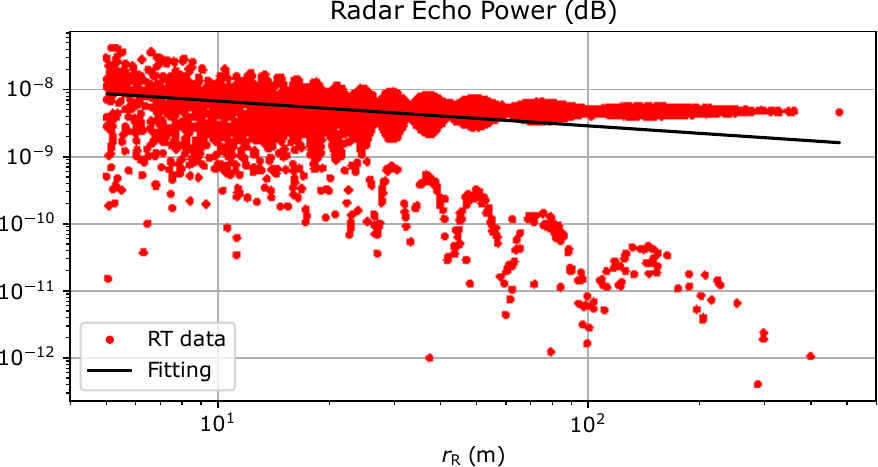}
    \caption{Without filtering}
    \label{fig:RT_power_R_clutter}
\end{subfigure}
\vspace*{0.1cm}

\begin{subfigure}{\linewidth}
    \centering
    \includegraphics[width=\linewidth]{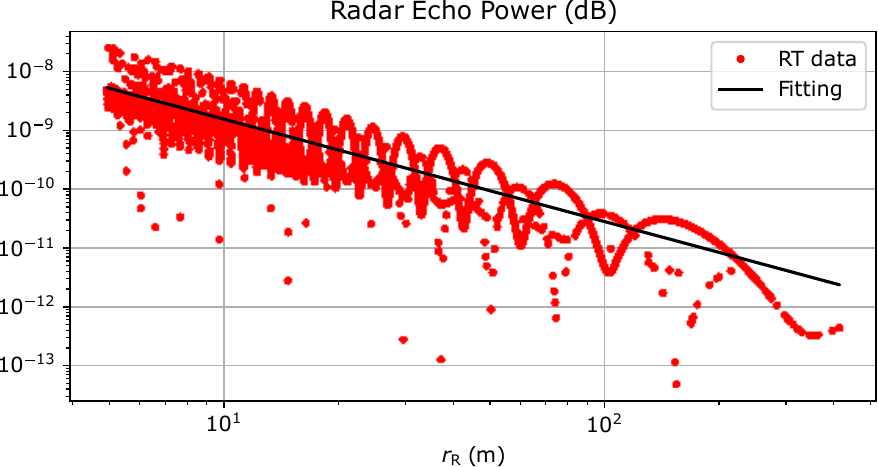}
    \caption{With filtering}
    \label{fig:RT_power_R_no_clutter}
\end{subfigure}
\caption{Received radar echo power with RT simulations, and fitting, with and without the clutter filtered out. The distance $r_\R$ denotes the distance between the radar antennas and the next vehicle ahead.}
\label{fig:RT_power_R}
\end{figure}

The parameters of the system models are extracted from the generated RT dataset. In order to perform this estimation, the different nodes densities are first estimated by averaging the number of nodes over the different network realisations. The densities extracted from the dataset are equal to $18.6$ km$^{-1}$ for the vehicles, and $1.79$ km$^{-1}$ for the interfering nodes. These lower densities are consequence of the removal of overlapping vehicles.\smallskip

Then, for the path-loss parameters, Least Square Estimation (LSE) in the log-log domain is applied. More specifically, denoting as $S^\text{RT}_{v,n}$ with $v\in\{\R,\C,\I\}$ the received power for each link of the $n^{\text{th}}$ simulation, and $r_{v,n}$ the associated distance, the LSE problem is written as 
\begin{equation}
\hat{\alpha}_v, \hat{\beta}_v = \arg\min_{\alpha_v,\beta_v} \sum_{n=0}^{N-1} \left(\left[S^\text{RT}_{v,n}\right]_\text{dB} - \left[\beta_v \right]_\text{dB} - \alpha_v \left[r_n\right]_\text{dB}\right)^2,
\end{equation}
where the operator $\left[\cdot\right]_\text{dB}$ denotes the conversion in dB scale. Note that, for the interference, the power $S_{v,n}^\text{RT}$ stands for the power of the interference coming from one single vehicle. The total interference is obtained by summing the power from each interfering vehicle within a simulated scenario. Furthermore, the LSE can be applied after separating the LoS and NLoS links, or mixing all the links together. The estimated parameters are summarised in Table \ref{tab:RT_prop_params}. Low path-loss exponents are estimated for the radar and communication links, owing to the canyon-like structure of the street, even with the presence of perpendicular streets. Contrariwise, the path-loss exponent of the interfering link is very high, owing to the strong blockage and high NLoS probability induced by the vehicles driving on both lanes. For the radar link, two cases are illustrated in Figure \ref{fig:RT_power_R}: either the echoes from the environment are kept, or they are filtered out. The echoes from the environment comprise all the rays which does not interact with the next vehicle ahead. This encompasses diffraction on building edges, or reflection and diffraction on other vehicles. Since these rays do not interact with the next vehicle ahead, they correspond to the high power data points in Figure \ref{fig:RT_power_R_clutter} which do not decrease with the distance. Cancelling these echoes leads to Figure \ref{fig:RT_power_R_no_clutter}. Oscillations of the power with the distance are explained by diffractions on the edges of the vehicle, with or without reflections on the buildings. Instead of using a linear path-loss model, these may be modelled more accurately, for instance following the approach of \cite{5645701}. Yet, this is not considered in this work, as such fitting does not improve significantly the results of the comparisons performed in Section \ref{sec:numerical_analysis}. \smallskip

Regarding the LoS probability model, the distances $d_{1,v}$ for $v \in\{\C,\I\}$ are set equal to the minimum distances $r_{v \text{min}}$ of the communication and interfering links. Yet, the distances $d_{2,v}$ are instead computed by solving another LSE problem: 
\begin{equation}
\hat{d}_{2,v} = \arg\min_{d_{2,v}} \sum_{n=0}^{N-1} \left(p^\text{RT}_{\text{L},v,n} - p_{\text{L},v}(r_{v,n};d_{1,v},d_{2,v})\right)^2,
\end{equation}
where $p^\text{RT}_{\text{L},v,n}$ are the LoS probabilities of the communication and interfering links estimated from the $n^\text{th}$ RT simulation, associated with the distance $r_{v,n}$, and $p_{\text{L},v}$  is the LoS probability model of \eqref{eq:LoS_prob}. This fitting leads to $\hat{d}_{2,\C} = 110.44$ m for the communication link, and $\hat{d}_{2,\I} = 43.95$ m for the interfering links. Indeed, the estimated distance is shorter for the interfering link, as the interfering signals encounters vehicles driving in both lanes, which increases the blockage probability.\smallskip

Finally, for the small-scale fadings, the $K$-factors are estimated by computing the ratio between the power from the RT outputs, and the estimated path-loss models at the same distance. This operation has shown that the small-scale fading of the three links can be modelled with Rayleigh fading to achieve a better fitting.

%% file: numerical_analysis.tex
\begin{figure}
    \centering
    \includegraphics[width=\linewidth]{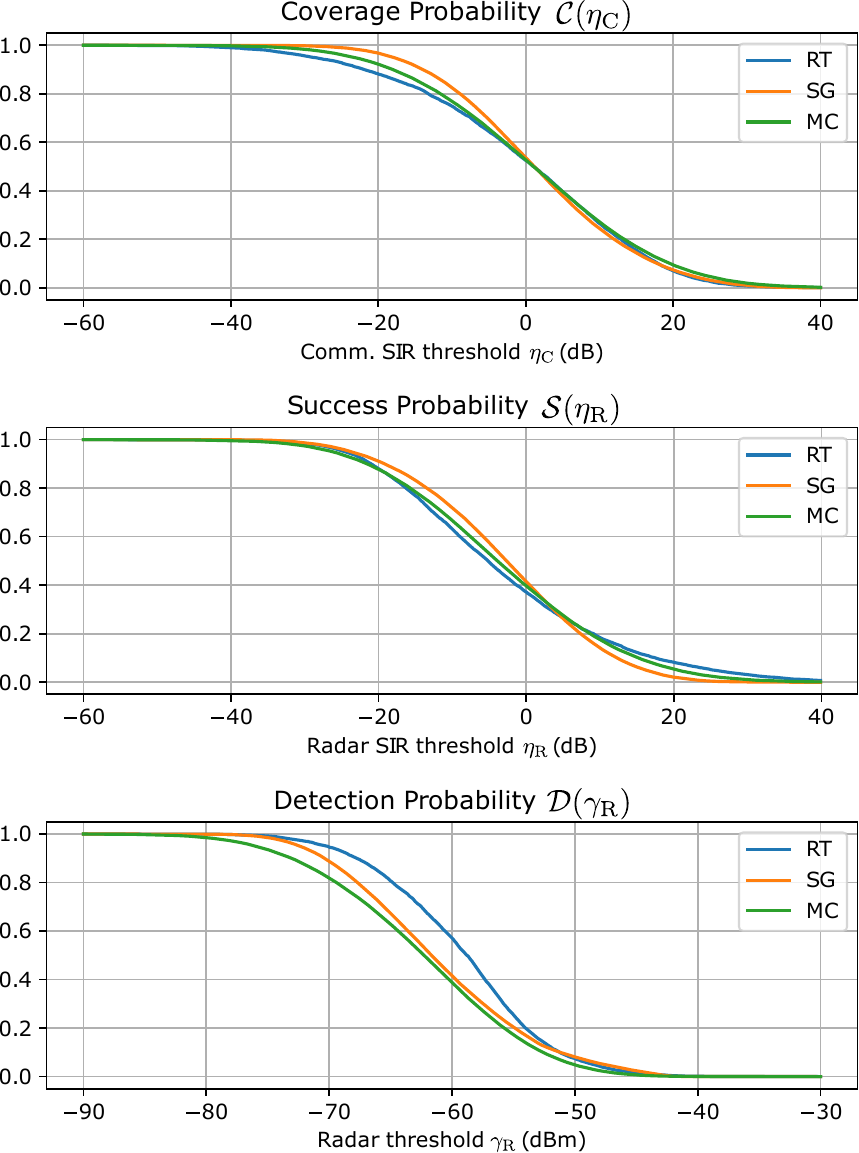}
    \caption{Comparison of the univariate metrics obtained with the SG, MC, and RT frameworks.}
    \label{fig:RT_univariate}
\end{figure}

\begin{figure*}
    \centering
    \includegraphics[width=\linewidth]{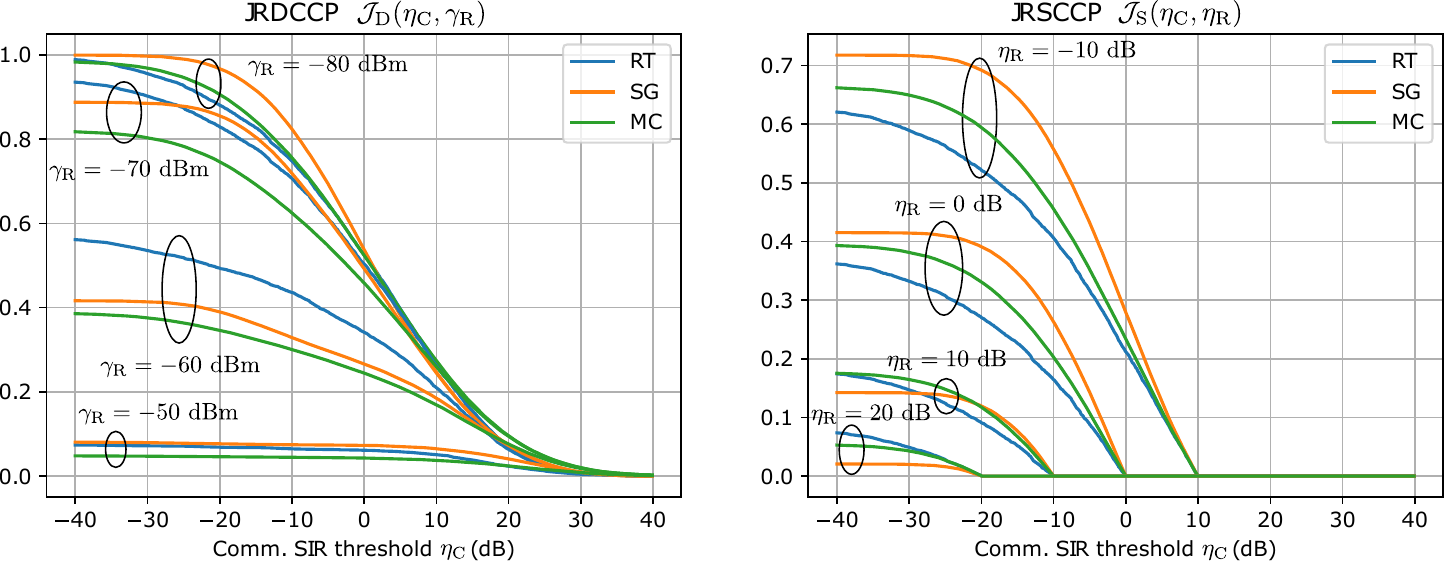}
    \caption{Comparison of the joint metrics obtained with the SG, MC, and RT frameworks.}
    \label{fig:RT_bivariate}
\end{figure*}

In this section, the metrics evaluated within the SG framework, RT framework (Section \ref{sec:RT_system_model}), and MC framework (Section \ref{sec:MC_system_model}) are compared, using the parameters obtained in Section \ref{sec:parameters_fitting}. First, Figure \ref{fig:RT_univariate} represents the univariate metrics, i.e. the coverage, success and detection probabilities. For the coverage and success probabilities, the metrics obtained with the RT and MC models are very close to each other, whereas the SG slightly overestimate the slope. Nonetheless, for the detection probability, larger differences are observed. Contrariwise to the other metrics, the MC gives lower detection probabilities.\smallskip

Then, Figure \ref{fig:RT_bivariate} represents the joint metrics, namely the JRDCCP and JRSCCP. Unfortunately, large discrepancies are observed between the three framework. Still, for the JRSCCP, the MC system model seems closer to the RT results. The main difference is the blockage, which is not modelled at all in the SG framework, and not well modelled in the MC framework through the LoS probability function. Consequently, the modelling of the interfering power is less accurate in the SG and MC frameworks, impacting both the radar and communication functions in the joint metrics evaluation.

%% file: conclusion.tex
In this paper, the performance evaluation in an ISAC V2X scenario is considered, within the SG, MC, and RT frameworks. After performing a fitting of the densities and propagation parameters, it has been shown that the SG and MC system models are relevant w.r.t. RT simulations for the evaluation of univariate communication and sensing metrics. Nonetheless, for the joint metrics which have been introduced, the results obtained with these two system models are rather inaccurate owing to the interference modelling, but similar trends are still observed. In future works, the model mismatches between the frameworks must be further investigated. For instance, it may be possible to slightly modify the SG model to take into account the blockage of other vehicles, or diffraction mechanisms, in order to improve the accuracy for joint ISAC metrics, with an acceptable increase of complexity.